\documentclass[sigconf]{acmart}
\usepackage{algorithmic}
\usepackage{graphicx}
\usepackage{textcomp}
\usepackage{xcolor}
\usepackage{xspace}
\usepackage{pifont}
\usepackage{subfigure}
\usepackage{multirow}
\usepackage{makecell}
\usepackage{booktabs}
\usepackage{array}
\newcolumntype{C}[1]{>{\centering\arraybackslash}m{#1}}
\newcommand{\acronym}{\textit{SLED}\xspace}
\renewcommand{\shortauthors}{Li et al.}
\AtBeginDocument{%
  }

\setcopyright{acmlicensed}
\copyrightyear{2025}
\acmYear{2025}
\acmDOI{XXXXXXX.XXXXXXX}
\acmConference[SEC 2025]{ACM/IEEE Symposium on Edge Computing}{December 03--06,
  2025}{Washington, D.C., USA}
\acmISBN{978-1-4503-XXXX-X/2018/06}

\begin{document}

\title{\acronym: A Speculative LLM Decoding Framework for Efficient Edge Serving}

\author{Xiangchen Li}
\orcid{0000-0002-5293-5348}
\affiliation{%
  \institution{Virginia Tech}
  \city{Blacksburg}
  \state{Virginia}
  \country{USA}
}
\email{lixiangchen@vt.edu}

\author{Dimitris Spatharakis}
\affiliation{%
  \institution{National Technical University of Athens}
  \city{Athens}
  \country{Greece}
}
\email{dspatharakis@netmode.ntua.gr}

\author{Saeid Ghafouri}
\affiliation{%
 \institution{Queen's University Belfast}
 \city{Belfast}
 \state{Northern Ireland}
 \country{UK}}
\email{s.ghafouri@qub.ac.uk}

\author{Jiakun Fan}
\affiliation{%
  \institution{Virginia Tech}
  \city{Blacksburg}
  \state{Virginia}
  \country{USA}}
\email{jiakunfan@vt.edu}

\author{Hans Vandierendonck}
\affiliation{%
  \institution{Queen's University Belfast}
  \city{Belfast}
  \state{Northern Ireland}
  \country{UK}}
\email{h.vandierendonck@qub.ac.uk}

\author{Deepu John}
\affiliation{%
  \institution{University College Dublin}
  \city{Dublin}
  \country{Ireland}}
\email{deepu.john@ucd.ie}

\author{Bo Ji}
\affiliation{%
  \institution{Virginia Tech}
  \city{Blacksburg}
  \state{Virginia}
  \country{USA}}
\email{boji@vt.edu}

\author{Dimitrios~S. Nikolopoulos}
\affiliation{%
  \institution{Virginia Tech}
  \city{Blacksburg}
  \state{Virginia}
  \country{USA}}
\email{dsn@vt.edu}

\renewcommand{\shortauthors}{Li et al.}
\begin{abstract}
The growing gap between the increasing complexity of large language models (LLMs) and the limited computational budgets of edge devices poses a key challenge for efficient on-device inference, despite gradual improvements in hardware capabilities. Existing strategies, such as aggressive quantization, pruning, or remote inference, trade accuracy for efficiency or lead to substantial cost burdens. This position paper introduces a new framework that leverages speculative decoding, previously viewed primarily as a decoding acceleration technique for autoregressive generation of LLMs, as a promising approach specifically adapted for edge computing by orchestrating computation across heterogeneous devices. We propose \acronym, a framework that allows lightweight edge devices to draft multiple candidate tokens locally using diverse draft models, while a single, shared edge server verifies the tokens utilizing a more precise target model. To further increase the efficiency of verification, the edge server batches the diverse verification requests from devices. This approach supports heterogeneous devices and reduces server-side memory footprint by sharing a single upstream target model across devices. Our initial experiments with Jetson Orin Nano, Raspberry Pi 4B/5, and an edge server equipped with 4 Nvidia A100 GPUs indicate substantial benefits: ×2.2 higher system throughput, ×2.8 higher system capacity, and better cost efficiency, all without sacrificing model accuracy.
\end{abstract}




\acmYear{2025}\copyrightyear{2025}
\setcopyright{cc}
\setcctype[4.0]{by-nc}
\acmConference[SEC '25]{The Tenth ACM/IEEE Symposium on Edge Computing}{December 3--6, 2025}{Arlington, VA, USA}
\acmBooktitle{The Tenth ACM/IEEE Symposium on Edge Computing (SEC '25), December 3--6, 2025, Arlington, VA, USA}
\acmDOI{10.1145/3769102.3770608}
\acmISBN{979-8-4007-2238-7/25/12}

\maketitle

\section{Introduction}
LLMs have revolutionized various domains, demonstrating remarkable capabilities in natural language understanding, generation, and complex reasoning \cite{NEURIPS2020_1457c0d6}. Their widespread adoption has led to transformative applications in areas such as intelligent chatbots, content creation, code generation, and scientific discovery. However, the immense memory and compute footprint associated with state-of-the-art LLMs, often comprising billions or even trillions of parameters, pose significant challenges for deployment. These models typically demand powerful accelerators like GPUs and substantial memory, limiting their direct execution on resource-constrained devices.
Deploying LLMs at the edge, closer to data sources and end-users, offers significant advantages including reduced latency, enhanced privacy, and lower bandwidth consumption\cite{8763885}. Nevertheless, edge environments, characterized by limited memory, processing power, and energy budgets, present formidable obstacles to efficient LLM inference. Existing strategies to address these limitations include aggressive model compression techniques such as quantization \cite{MLSYS2024_42a452cb, zeng2024abqllmarbitrarybitquantizedinference}, pruning\cite{ma2023llmprunerstructuralpruninglarge,sun2024simpleeffectivepruningapproach}, and knowledge distillation \cite{hinton2015distillingknowledgeneuralnetwork}. Other approaches involve distributed inference, where model layers are partitioned across multiple devices or between edge and cloud\cite{10818760,ye2025jupiter}, or full remote inference, where the entire computation is offloaded to a powerful central server\cite{gao2025collaborativespeculativeinferenceefficient,yu2024edgellmenablingefficientlarge}. While these methods show some potential, they often come with trade-offs: compression can sacrifice model accuracy, distributed inference introduces synchronization overheads and is incompatible with heterogeneous edge devices, and remote inference negates the benefits of edge deployment, incurring non-negligible costs.

\begin{figure}[!t]
    \centering
    \includegraphics[scale=0.6]{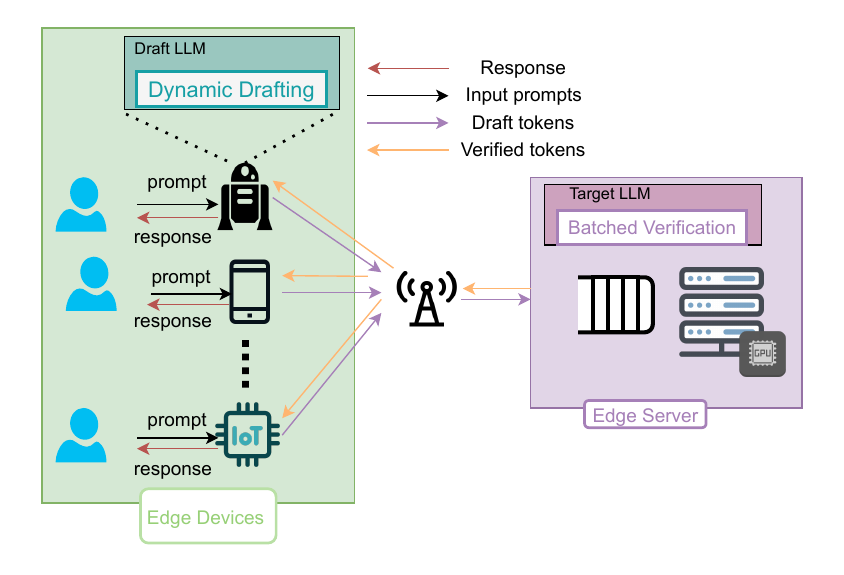}
    \caption{System overview of the proposed speculative LLM decoding framework for efficient edge serving.}
    \Description{Overview of SLED: edge devices run a draft LLM to propose tokens, which are sent over the network to an edge server running a target LLM that verifies or rejects them; prompts come from users on devices and verified tokens return as responses.}
    \label{fig_system_overview}
\end{figure}

Speculative decoding \cite{leviathan2023fast} is a decoding acceleration technique that first generates multiple draft tokens using a relatively small draft model and then verifies them in a single forward pass using a larger, more accurate target model. By generating multiple tokens with the smaller model and validating them in a single pass on the larger model, speculative decoding significantly reduces the number of forward passes required on the large model, thereby accelerating the decoding process. This position paper introduces \acronym, shown in Fig. \ref{fig_system_overview}, a novel approach that re-imagines speculative decoding as a paradigm specifically tailored for efficient LLM inference at the edge, by intelligently orchestrating computation across heterogeneous devices. Within the defined service area consisting of an edge server and multiple heterogeneous edge devices, and each edge device is equipped with its own lightweight LLMs scaled according to individual computational and memory resource capacities. These edge devices are responsible for serving diverse LLM-based applications such as intelligent personal assistants, text generation, and semantic analysis, among other tasks. Concurrently, a single, shared edge server, equipped with a more precise target model, efficiently batches and verifies these drafted tokens. 

The advantages of the \acronym are threefold:
\begin{enumerate}
    \item Compared with inference solely on the edge device, the \acronym improves the quality of response on the device by leveraging a larger target model on the server to verify draft tokens.
    \item Compared with inference solely on the edge server, \acronym reduces the monetary cost for edge users by limiting their use of server resources, requiring only token verification rather than full generation.
    \item It utilizes the computational resource of edge server to verified batched draft tokens from devices, rather than generate all tokens solely, enabling edge server support more edge devices simultaneously.
\end{enumerate}

We compare the system capacity, the number of edge devices supported by the system, of \acronym and a centralized serving system with the same response rate but different device types. From Tab. \ref{tab_capacity}, we observe that compared with a centralized LLM serving system for the edge, the proposed \acronym is capable of increasing the system capacity by 2.6 to 2.9 times.

\begin{table}[!t]
\footnotesize
  \caption{\label{tab_capacity}Capacity of \acronym and centralized serving system (Capacity = maximum number of supported devices at target response rate (devices))}
  \centering
  \setlength{\tabcolsep}{2pt}
  \begin{tabular}{l C{1.6cm} C{1.6cm} C{1.6cm} C{1.8cm}}
  \toprule
  System     & RPi 4b (llama.cpp) & RPi 5 (llama.cpp) & Nvidia Jetson \\
  \midrule
    \textit{\textbf{\acronym}}   & 18.30 & 5.24 & 19.53 \\
    \textbf{Centralized serving}   & 7.05 & 1.83 & 7.06 \\   
    \textbf{Capacity improvement}   & ×2.60 & ×2.86 & ×2.77 \\ 
  \bottomrule
  \end{tabular}
\end{table}

Our key contributions are summarized as follows:
\begin{itemize}
    \item We propose \acronym, a novel speculative decoding framework specifically designed for heterogeneous edge computing environments \textcolor{black}{by batching the verification requests from varying draft models on the server}, enabling efficient LLM inference without accuracy degradation.

    \item In \acronym, we propose and deploy the dynamic drafting scheme on edge devices. By dynamically requesting for verification according to the confidence score of the draft model, the edge devices can avoid unnecessary verification, hence reducing the communication rounds and improving the utilization of the server.

    \item We demonstrate through preliminary evaluation the substantial benefits of \acronym in terms of ×2.2 more system throughput, ×2.8 more system capacity, and better cost-efficiency on diverse edge hardware.
\end{itemize}

The remainder of this paper is organized as follows: Section \ref{sec:related_work} reviews existing work in LLM inference for edge computing. Section \ref{sec:sled} details the architectural design and key components of \acronym. Section \ref{sec:experiments} presents our experimental setup and discusses the evaluation results. Finally, Section \ref{sec:conclusion} concludes the paper and outlines future research directions.

\section{Related Work}\label{sec:related_work}

\begin{table}[!t]
\footnotesize
  \caption{\label{tab:comparison}Comparison of related work; Edge-Serving: Does the system support edge computing?; Heterogeneity: Is the heterogenity of edge devices considered in the system design?; Lossless: Whether does the system deliver LLM service without any performance degradation? Scalable Model: Is the system capable of scaling model according to conditions without too much overhead?}
  \centering
  \setlength{\tabcolsep}{2pt}
  \begin{tabular}[]{ccccc}
  \toprule
  System                                                        & Edge-Serving & Heterogeneity & Lossless & Scalable Model \\
  \midrule
    EdgeShard\cite{10818760}        & \textcolor{teal}{\ding{51}} & \textcolor{teal}{\ding{51}} & \textcolor{red}{\ding{53}} & \textcolor{red}{\ding{53}} \\
    Galaxy\cite{ye2025jupiter}      & \textcolor{teal}{\ding{51}} & \textcolor{red}{\ding{53}} & \textcolor{teal}{\ding{51}} & \textcolor{red}{\ding{53}} \\
    Orca\cite{yu2022orca}           & \textcolor{red}{\ding{53}} & \textcolor{red}{\ding{53}} & \textcolor{teal}{\ding{51}} & \textcolor{teal}{\ding{51}} \\
    vLLM\cite{kwon2023vllm}         & \textcolor{red}{\ding{53}} & \textcolor{red}{\ding{53}} & \textcolor{teal}{\ding{51}} & \textcolor{teal}{\ding{51}} \\
    FastServe\cite{wu2024fastserve} & \textcolor{red}{\ding{53}} & \textcolor{red}{\ding{53}} & \textcolor{teal}{\ding{51}} & \textcolor{teal}{\ding{51}} \\
    AWQ\cite{MLSYS2024_42a452cb}    & \textcolor{teal}{\ding{51}} & \textcolor{teal}{\ding{51}} & \textcolor{red}{\ding{53}} & \textcolor{red}{\ding{53}} \\
    MobileBERT\cite{sun2020mobile}  & \textcolor{teal}{\ding{51}} & \textcolor{teal}{\ding{51}} & \textcolor{red}{\ding{53}} & \textcolor{red}{\ding{53}} \\
    \acronym                        & \textcolor{teal}{\ding{51}} & \textcolor{teal}{\ding{51}} & \textcolor{teal}{\ding{51}} & \textcolor{teal}{\ding{51}} \\
    
  \bottomrule
  \end{tabular}
\end{table}

The efficient inference of LLMs on resource-constrained devices has been a focal point of research, broadly categorized into model compression techniques, distributed inference strategies, and remote offloading paradigms.

\subsection{Model Compression and Lightweight Architectures}
To enable LLMs to run on resource-constrained devices, significant efforts have been directed towards model compression. Quantization reduces the numerical precision of model parameters and activations to decrease memory footprint and accelerate computation \cite{dettmers2022llmint88bitmatrixmultiplication, Liu_2023, MLSYS2024_42a452cb}. Pruning identifies and removes redundant connections or neurons from the neural network without significant performance loss, resulting in sparser and smaller models~\cite{han2016deepcompressioncompressingdeep}. Knowledge distillation involves training a smaller "student" model to mimic the behavior of a larger "teacher" model, thereby transferring knowledge and achieving comparable performance with a significantly smaller footprint \cite{hinton2015distillingknowledgeneuralnetwork}. Beyond these optimization techniques, research has also focused on designing inherently lightweight transformer architectures that are more efficient from the ground up, such as MobileBERT \cite{sun2020mobile}, Mamba~\cite{gu2024mambalineartimesequencemodeling} or other compact variants, often by optimizing attention mechanisms or reducing the number of layers and hidden dimensions. \textcolor{black}{Recent efforts, such as EdgeLLM \cite{10812936}, have explored speculative decoding for on-device inference by introducing compute-efficient branch navigation and adaptive fallback strategies to reduce resource demands. However, co-locating both draft and target models on edge devices further strains limited resources, and the output quality remains constrained by the size of the deployable target model.} Despite their advantages in reducing model size and computational demands, a common limitation of these model compression techniques is the inherent trade-off with model quality: aggressive compression often leads to a measurable decrease in accuracy compared to their full-sized counterparts.

\subsection{Edge-Cloud/Server Offloading and Distributed Inference}
Another line of research focuses on distributing LLM computation across multiple devices or partitioning tasks between edge and cloud/server infrastructure. Model partitioning schemes divide a large LLM into smaller sub-models, with different parts executed on different devices \cite{10818760,ye2024galaxyresourceefficientcollaborativeedge,ye2025jupiter}. For example, EdgeShard\cite{10818760} partitioned LLM into shards and deploy on distributed devices to benefit from the collaboration among edge devices and cloud server. This often involves pipeline parallelism or tensor parallelism techniques, where different stages or segments of the model's computation are assigned to different devices. While this allows larger models to run on resource-constrained setups, it introduces communication overheads and synchronization challenges, particularly for heterogeneous hardware and varying network conditions. Edge-cloud offloading dynamically decides which parts of the inference task should be performed locally at the edge and which should be offloaded to a more powerful cloud server, often based on real-time resource availability, network bandwidth, and latency requirements \cite{10818760}. These methods aim to balance the benefits of edge processing with the computational power of the cloud, but often require sophisticated orchestration and robust connectivity.

\subsection{Pure Remote Inference}
Pure remote inference, where the entire LLM resides on centralized data-center GPUs, represents a prevalent deployment paradigm due to its simplicity and centralized resource utilization. Recent research primarily focuses on resource efficiency and latency optimization. Kwon et al.\cite{kwon2023vllm} proposed vLLM with a PagedAttention allocator, significantly reducing KV-cache overhead and fragmentation, achieving up to 4× throughput improvement. Wu et al. \cite{wu2024fastserve} introduced FastServe, leveraging multi-level feedback queue scheduling and proactive KV-cache management to reduce tail latency by up to 31× at the 99th percentile. Rajbhandari et al. \cite{rajbhandari2022deepspeed} developed DeepSpeed Inference, combining multiple parallelism strategies with NVMe and CPU off-loading, allowing inference of substantially larger models and reducing latency by up to 7.3×. Crucially, the standard decoding process in remote inference is often autoregressive, generating one token at a time, which can be memory-intensive due to large key-value caches and lead to resource under-utilization on powerful servers. Moreover, the cost associated with cloud GPU instances for continuous, often under-utilized, inference also presents a substantial economic burden, especially for high-throughput scenarios. 

Table \ref{tab:comparison} compares the \acronym and related works that deliver LLM inference service or propose model variants for edge devices, among which \acronym stands out as the only approach that enables lossless LLM inference for heterogeneous edge devices, while maintaining a collaborative design that can flexibly accommodate increasingly large models. \acronym directly addresses these limitations by fundamentally altering the inference paradigm. Instead of autoregressive token generation on the powerful central server, \acronym offloads the preliminary token drafting to lightweight edge devices. This allows the central server to focus its considerable resources primarily on the more efficient and batchable task of verifying multiple drafted tokens. By doing so, \acronym significantly improves the utilization of expensive server-side GPU resources without sacrificing model accuracy, leading to a more cost-effective and scalable distributed LLM inference system.

\section{\acronym Design} \label{sec:sled}

\begin{figure*}[t!]
    \centering
    \includegraphics[scale=0.45]{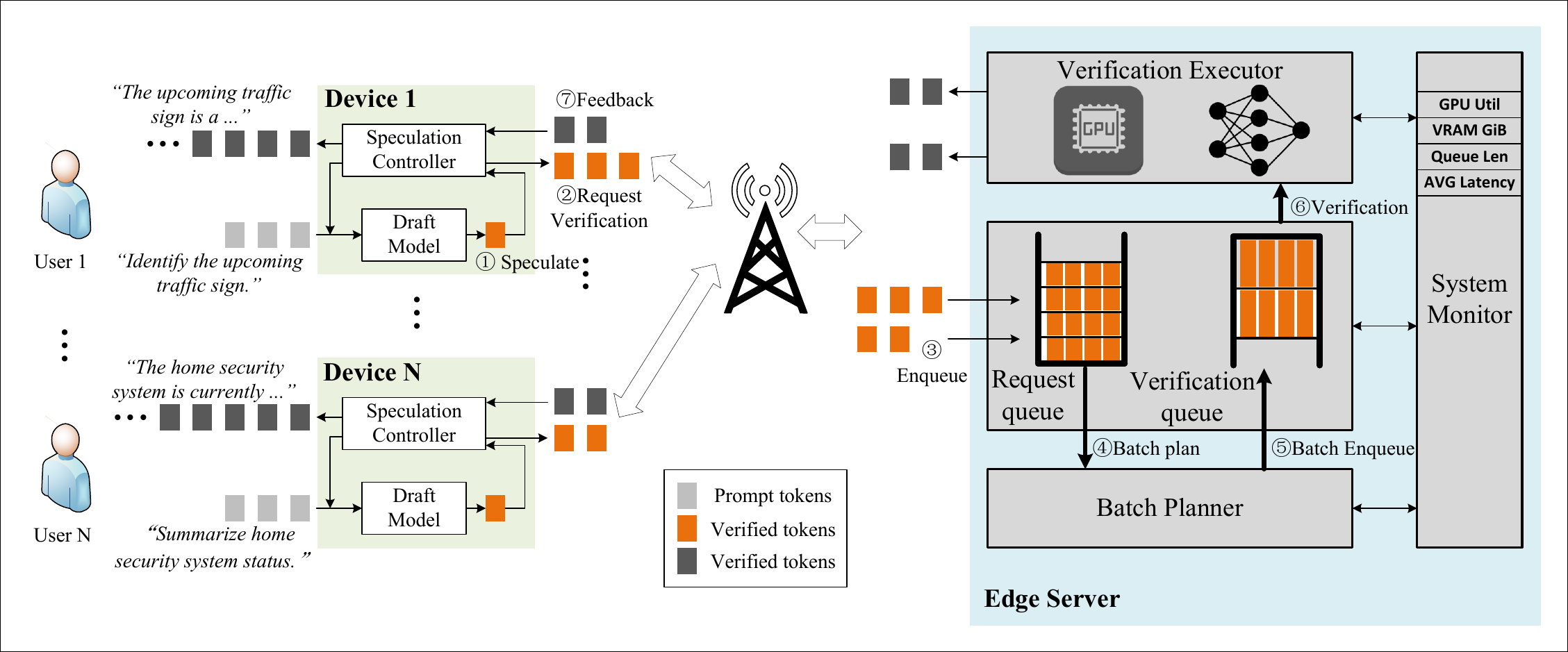}
    \caption{SLED architecture and request/verification data flow}
    \Description{Detailed SLED architecture and data flow: each device has a speculation controller and draft model; requests are enqueued on the server, a batch planner forms verification batches, a verification executor checks tokens, and feedback is returned to devices.}
    \label{fig_data_flow_overview}
\end{figure*}

\subsection{Speculative Decoding}
\textcolor{black}{\acronym adopts speculative decoding\cite{leviathan2023fast}, a recently proposed paradigm that leverages a lightweight draft model to generate multiple tokens. Instead of waiting for a single token from the large target model at each step, the draft model speculatively predicts a candidate sequence, which is then selectively verified or corrected by the more accurate but slower target model with one single forwarding. This design decouples the fast-path token generation from the slow-path verification, enabling improved throughput and reduced end-to-end latency without compromising output quality. In \acronym, this speculative execution is further optimized for edge-server deployment under variable connectivity and resource constraints.}

\textcolor{black}{According to the original work on speculative decoding\cite{leviathan2023fast}, a draft token  $\tilde{x}_{n+1}$ is sampled from the draft model’s output distribution $p(x \mid x_1, x_2, \cdots, x_n)$. This draft token is then accepted with probability:}
    
\textcolor{black}{
\begin{equation}
    \alpha = \min\!\left(1, \frac{p_{\mathrm{target}}(\tilde{x}_{n+1}\mid x_1,\dots,x_n)}{p_{\mathrm{draft}}(\tilde{x}_{n+1}\mid x_1,\dots,x_n)}\right),
\label{eq_sampling}
\end{equation}
}
\textcolor{black}{where $p_{\text{target}}$is the distribution given by the large target model, and $p_{\text{draft}}$ is that of the smaller draft model. Tokens that fail this acceptance criterion are resampled directly from $p_{\text{target}}-p_\text{{draft}}$. As a result, the generated sequence strictly follows the target model’s distribution, ensuring no degradation in accuracy compared with standard autoregressive decoding.}

\subsection{System Overview}

Fig. \ref{fig_data_flow_overview} shows the detailed structure and data flow of the \acronym. In close proximity to $N$ edge devices, typically located at facilities such as base stations, the edge server provides substantial computational capabilities, leveraging specialized hardware like Graphics Processing Units (GPUs) or Neural Processing Units (NPUs). On this edge server, a single, comprehensive target model is deployed, optimized for efficiently verifying the draft tokens generated by the distributed edge devices.

Operationally, user-generated prompts, encompassing a wide array of task-specific requests, are initially received and tokenized locally by each edge device. Subsequently, the tokenized prompts, denoted as input sequences $p^n$ where $n \in \{1, 2, \ldots, N\}$, are processed by local draft models to generate speculative tokens. These draft tokens are then transmitted to the edge server for verification. Upon completion of the verification step, the edge server communicates the results back to the respective edge devices, specifically identifying rejected token positions along with any necessary corrective tokens.

This drafting-verification workflow iteratively progresses, alternating between local speculation at the edge devices and centralized validation at the edge server, until the generated output reaches the predetermined desired length or the end-of-response token is encountered. This collaborative mechanism not only optimizes resource utilization by distributing computational tasks according to device capabilities but also significantly reduces latency and enhances overall efficiency by transmitting tokens rather than huge activations.

\subsection{Dynamic Drafting on Edge Devices}

On edge devices, each verification cycle is preceded by the generation of multiple draft tokens. The acceptance rate of these draft tokens serves as a crucial indicator of their quality, directly influenced by the capabilities of the draft models utilized. A higher acceptance rate is desirable as it signifies fewer verification iterations and consequently reduces the computational burden on the costly target model, thereby mitigating communication overhead inherent in edge computing scenarios.

\begin{figure}[!t]
    \centering
    \includegraphics[scale=0.3]{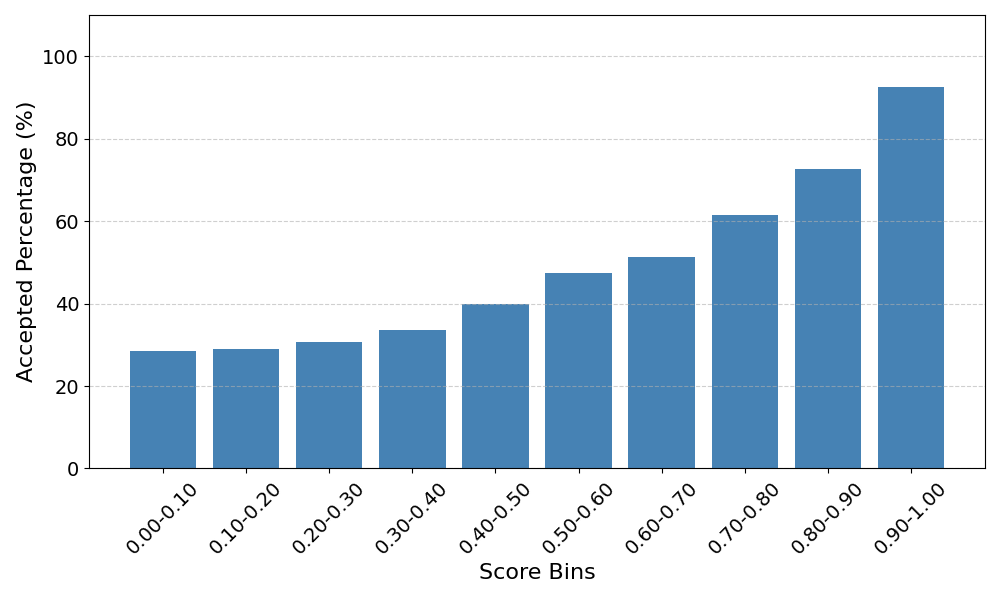}
    \caption{Acceptance rate vs. confidence of draft tokens}
    \Description{Bar chart showing acceptance percentage increases monotonically with the confidence score of draft tokens; higher confidence bins lead to more tokens being accepted.}
    \label{fig_confidence_vs_acceptance}
\end{figure}

Previous studies \cite{huang2025specserve}, validated by our preliminary experimental results, have established a correlation between the acceptance rate of draft tokens and their associated confidence scores derived from the output logits. As illustrated in Fig.\ref{fig_confidence_vs_acceptance}, draft tokens with higher confidence scores exhibit a significantly increased likelihood of acceptance by the target model. 

Building upon this insight, we propose and implement a dynamic drafting mechanism on edge devices. This adaptive strategy modulates the speculative decoding length based on the real-time evaluation of token confidence scores. Formally, we introduce a threshold parameter, $c_{th}$, derived empirically, and define the decision-making process for triggering server verification for the draft tokens as follows:

\begin{equation}
\label{eq1}
    c_s^i 
    \begin{cases}
        < c_{th}, request \;verification \\
        \geq c_{th}, generate\; another\; token \\
    \end{cases},
\end{equation}
where $c_s^i$ represents the confidence score associated with token $t_s^i$.

Considering the inherent unreliability and fluctuating nature of network conditions in edge computing environments, such as variable round-trip time (RTT) and intermittent connectivity, we further enhance our system with an asynchronous decoding mechanism accompanied by a timeout protocol. Specifically, edge devices continue generating additional draft tokens using local lightweight LLM concurrently while awaiting verification responses from the edge server. If a verification response confirms acceptance of all previously sent draft tokens, these locally generated tokens seamlessly transition into the draft token queue for subsequent verification cycles, thus significantly reducing idle wait times.

Additionally, each verification request initiates a timer on the device side. If the verification response exceeds the timer due to server failures or network disruptions, the most recently-produced draft tokens are concatenated with existing draft tokens for subsequent verification attempts. To maintain continuity of user experience, the draft tokens generated during this period are released to users as a fallback when consecutive verification failures exceed the threshold.

\subsection{Batched Verification on Edge Server}

The edge server aggregates verification requests from multiple edge devices into batches to optimize computational efficiency and throughput. Our current implementation within \acronym{} employs a static batching strategy. Under this scheme, incoming verification requests are temporarily queued until reaching a fixed batch size. Subsequently, a batch planner retrieves the queued requests, applies appropriate padding to equalize token lengths, and forwards the consolidated batch to the target LLM for verification.

A critical advantage of \acronym lies in the target model’s ability to accept and verify draft tokens generated by diverse draft LLMs across heterogeneous edge devices \textcolor{black}{provided that all edge devices share the same tokenizer}. This compatibility effectively mitigates device heterogeneity, enabling each device to select a draft model suited to its computational constraints while ensuring scalable and efficient inference across a wide range of edge hardware.

\section{Evaluation}\label{sec:experiments}
In this section, we evaluate the performance and efficiency of the proposed \acronym framework through extensive simulations and measurements. We assess \acronym's efficacy compared to a centralized LLM serving system which serves the decoding requests from edge device directly, and the edge-only inference system which generates all tokens locally on the devices across various metrics including throughput, system capacity, cost efficiency, and impact of speculative length on system capacity and the throughput. 

To accurately simulate verification request workloads from edge devices utilizing speculative decoding, we adopt a Poisson-based modeling approach. Each edge device is considered an independent source of verification requests, with inter-arrival times following an exponential distribution. This modeling choice effectively captures the asynchronous and inherently stochastic nature of real-world device interactions, ensuring that the simulated workload closely mirrors realistic operational conditions. The device-specific request rate is derived directly from realistic device speculative decoding throughput, ensuring that the simulation's temporal patterns closely align with practical speculative decoding workloads.


As for the device setting, we tested Raspberry Pi 4b, 5 and NVIDIA Jetson Orin Nano on the \acronym system supported by 4 A100 GPUs. \textcolor{black}{We deployed LLaMA family, including model sizes of 1B, 3B, 11B and 70B on the edge devices and verification server, supporting up to context length of 128k.}

\subsection{Whole System Token Generation Rate (WSTGR)}
\begin{figure}[t]
    \centering
    \includegraphics[scale=0.43]{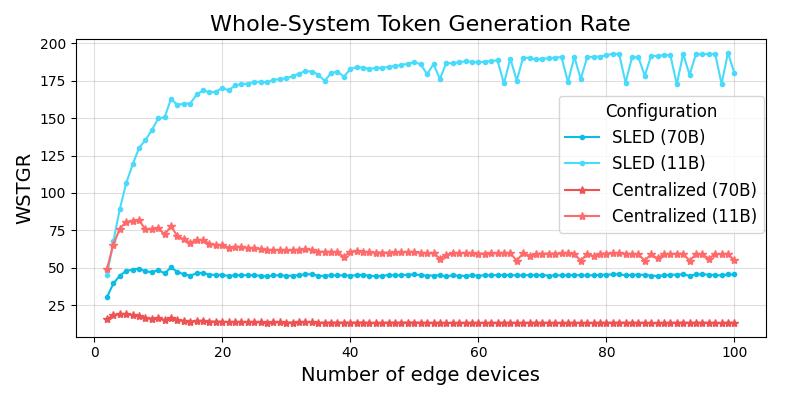}
    \caption{WSTGR comparison between \acronym and centralized LLM serving systems, highlighting improved scalability of the \acronym framework. }
    \Description{Line plot of whole-system token generation rate versus number of edge devices comparing SLED and a centralized serving baseline; SLED curves rise and scale with more devices while centralized baselines stay much lower.}
    \label{fig_wstgr}
\end{figure}
We first evaluate the Whole System Token Generation Rate (WSTGR), which is defined as the total number of tokens generated and verified by the entire inference system per second, and serves as a metric for the system’s overall productive output\cite{liu2024optimizing}. Given a certain time period we measure the total number of tokens generated by the \acronym and centralized inference system. The verification workload model is derived from a Raspberry Pi 5 device running a LLaMA 3B model. Additionally, we evaluate two different target models (11B and 70B) on both the \acronym system and a centralized inference system. As shown in Fig. \ref{fig_wstgr}, for both the 11B and 70B models, the WSTGR increases rapidly in the initial stages as batch size grows, due to the amortization of fixed GPU launch and driver overhead, and improved utilization of GPU cores. The proposed \acronym system achieves higher overall throughput than the centralized serving system under identical conditions, including the same number of devices and target model. This observed scalability demonstrates that \acronym effectively utilizes distributed edge resources to enhance the system’s token generation capacity.

The more than twofold improvement in WSTGR over the centralized serving system stems from the efficient and balanced distribution of computational tasks in the \acronym system. Specifically, in \acronym, simple token generation tasks can be handled by relatively small models \cite{leviathan2023fast}, such as those deployed on edge devices, while more complex tasks are offloaded to larger models on the edge server. This architectural separation allows computation to be distributed across edge devices and the edge server in a more resource-aligned and efficient manner. As a result, the computational capacity of the edge server is reserved for challenging verification tasks, rather than being consumed by processing simpler tokens from edge devices, unlike in a centralized LLM serving setup.

\subsection{Speculative Length vs. throughput and capacity}
\begin{figure}[t]
    \centering
    \includegraphics[scale=0.45]{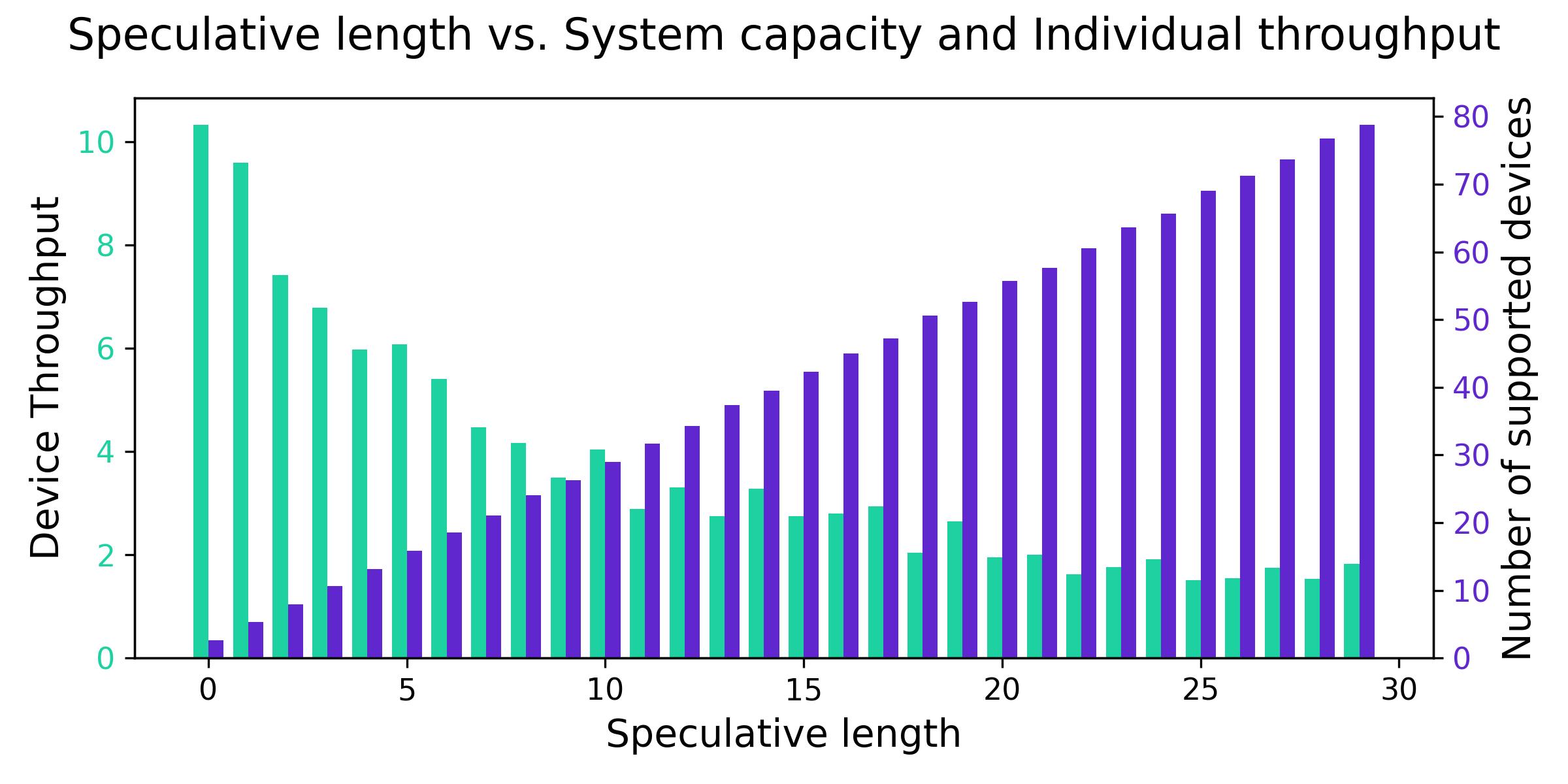}
    \caption{The Impact of speculative length on system-level capacity and device-level throughput, showing the speculative length should be considered to balance the tradeoff.}
    \Description{Grouped bar chart showing the trade-off as speculative length grows: device-level throughput decreases while system capacity—the number of supported devices—increases, indicating a balance point is needed.}
    \label{fig_wsgtr_speclen}
\end{figure}
In speculative decoding, the length of the draft sequence used for verification on the target model is defined as the speculative length, and it affects both token generation throughput and the capacity of \acronym, that is, the number of edge devices supported by \acronym simultaneously. In this experiment, we manually adjust the speculative length for drafting using LLaMA 1B model on a Raspberry Pi 5 device, and measure both the device throughput and overall system capacity. As shown in Fig. \ref{fig_wsgtr_speclen}, increasing the speculative length results in lower device throughput but higher system capacity. This inverse relationship between per-device and system-level metrics highlights the importance of selecting an appropriate speculative length to balance the performance of individual edge devices and the system as a whole. On one hand, a longer speculative length reduces token generation on each device, since the drafting throughput remains stable, and a longer speculative length leads to a longer verification period, thereby reducing the response update rate. On the other hand, a longer verification period for individual devices reduces the verification workload on the edge server, allowing it to support more devices concurrently.

\subsection{Cost Efficiency and System Throughput}
\begin{figure}[t]
    \centering
    \includegraphics[scale=0.38]{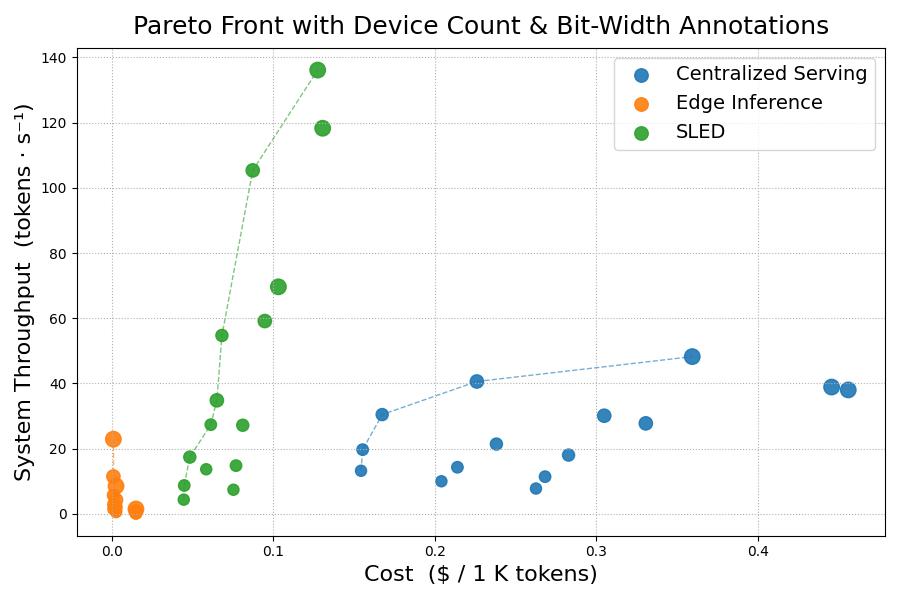}
    \caption{Pareto front showing optimal trade-offs between energy consumption per token and latency, highlighting the efficiency of \acronym.}
    \Description{Scatter plot Pareto front of system throughput versus cost per thousand tokens with point annotations; SLED points form a superior front, indicating better efficiency for given cost compared with centralized or pure edge inference.}
    \label{fig_pareto}
\end{figure}
The cost efficiency of token generation in edge computing scenarios is a critical factor and has been considered in various edge inference system designs \cite{Husom2025SustainableLLM, Jang2025EdgeFirst}. In this paper, we compare the proposed \acronym system, a centralized serving system, and an all-edge decoding system in terms of both cost efficiency and WSTGR. To systematically analyze the trade-off between cost and performance, we construct Pareto front visualizations, which highlight the non-dominated configurations that achieve the best balance between monetary cost and system throughput.

In our experiments, the cost and throughput metrics for edge inference scenarios were carefully computed based on a comprehensive capital expenditure (CAPEX) and operational expenditure (OPEX) model\cite{walker2009cpu}. Specifically, we adopt the widely recognized CPU-hour cost model described by Walker\cite{walker2009cpu} and the edge-compute modeling approach proposed by Eriksson\cite{eriksson2020cost}. The CAPEX component was determined by amortizing the purchase price of each edge device (Raspberry Pi 5 priced at \$80 \cite{upton2023pi5}) over a three-year lifetime, assuming an average device utilization rate of 70\%. The OPEX component included electrical consumption calculated from experimentally measured average power draw (8 W for Pi 5) \cite{odonnell2023review} and industrial electricity rates (0.083 \$/kWh) \cite{eia2025electric}. Combining these costs, we obtained a unified hourly expense for each device, subsequently normalized by the experimentally measured token generation rates (tokens per second), as shown in Eq. \ref{eq_cost}. The resulting metric, expressed clearly as dollars per one thousand generated tokens (\$/1K tokens), enabled direct and transparent comparison across different experimental configurations and devices.

\begin{equation}
    \text{Cost}\, =
\frac{1000}{3600\,R}\,
\Bigl(
  \frac{P_{\text{device}}}{3 \times 8760 \times 0.70}
  \;+\;
  \frac{P_{\text{avg}}}{1000}\times 0.083
\Bigr)
\label{eq_cost}
\end{equation}

Figure \ref{fig_pareto} compares the following three deployment strategies along a common cost–performance plane. Specifically, the strategies are: 1) all-Server executes every token‐generation step on a bank of four NVIDIA A100–80 GB GPUs. 2) All-Edge places the same LLaMA draft model on each Raspberry Pi 5, with no server involvement. 3) \acronym lets the Raspberry Pi 5 generate draft tokens, which are batch-verified on the A100 cluster with the same configuration of the centralized scenario. For every strategy, we sweep two orthogonal factors: quantization precision (16-, 8-, and 4-bit) and edge-device count $N\in\{1,2,4,8,16\}$. Cost is monetised as dollars per one-thousand verified tokens.

We observe that \acronym 's skyline consistently dominates the Pareto frontier, achieving lower cost per 1K verified tokens while sustaining higher overall throughput. For instance, with the same system capacity and quantization level, \acronym achieves a throughput of 137 tokens/s—representing a 3.5× improvement over the centralized baseline—while simultaneously reducing cost to just 29\% of that. This advantage becomes more pronounced as the number of edge devices increases. Furthermore, quantization universally improves cost efficiency across all schemes by simultaneously reducing energy demand and increasing per-device generation rate. Notably, the 4-bit \acronym configuration with 16 devices achieves 137 tokens/s at \$0.13 / 1K tokens, representing a 65\% improvement in throughput over the best-performing All-Edge setup, with acceptable additional cost. These results substantiate the claim that \acronym enables a superior cost–throughput trade-off, combining local cost efficiency with global throughput.

\subsection{Impacts of Network Connectivity}

\begin{figure}[t!]
    \centering
    \includegraphics[scale=0.21]{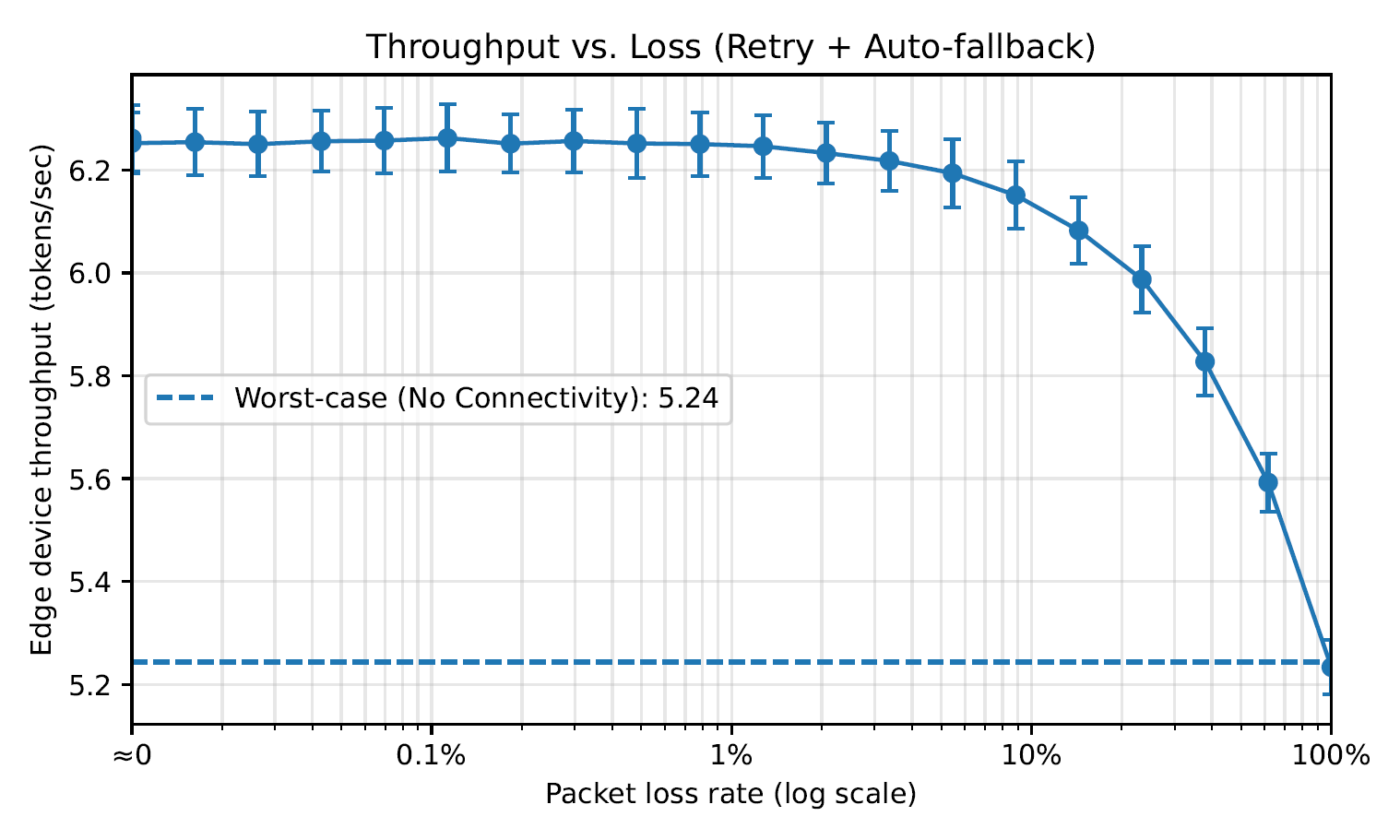}
    \caption{\textcolor{black}{The impact of packet loss on edge device-side throughput.}}
    \Description{Line plot of edge-device throughput versus packet loss rate on a log-scaled x-axis; throughput remains stable at low loss and then drops sharply beyond roughly ten percent as retry and fallback dominate.}
    \label{fig_packageloss_throughput}
\end{figure}

\textcolor{black}{We built a \acronym prototype to quantify how network loss affects edge-side throughput. The edge runs on a Raspberry Pi 5B hosting meta-llama/Llama-3.2-1B-Instruct for drafting, while the server uses A100 GPU running meta-llama/Meta-Llama-3.1-70B-Instruct for batched verification. We instrumented the system to sweep packet-loss ratios from 0\% to 100\% under a fixed RTT budget and measured the committed token throughput at the edge. As shown in Fig.\ref{fig_packageloss_throughput}, across commonly used settings in practice ($\leq$5–10\% loss), SLED’s edge throughput degraded only slightly (typically <2–3\% versus the no-loss baseline), confirming that the proposed retry-then-fallback policy effectively masks moderate loss. Even under complete network unavailability (responses never arrive), the edge maintained $\geq$5.24 tokens/s, comfortably above typical human reading rates, ensuring responsive user-perceived progress. These results indicate that, in both engineering practice and research deployments, SLED preserves acceptable edge throughput under realistic packet-loss regimes, while providing graceful performance under extreme conditions.}

\begin{figure}[t!]
    \centering
    \includegraphics[scale=0.52]{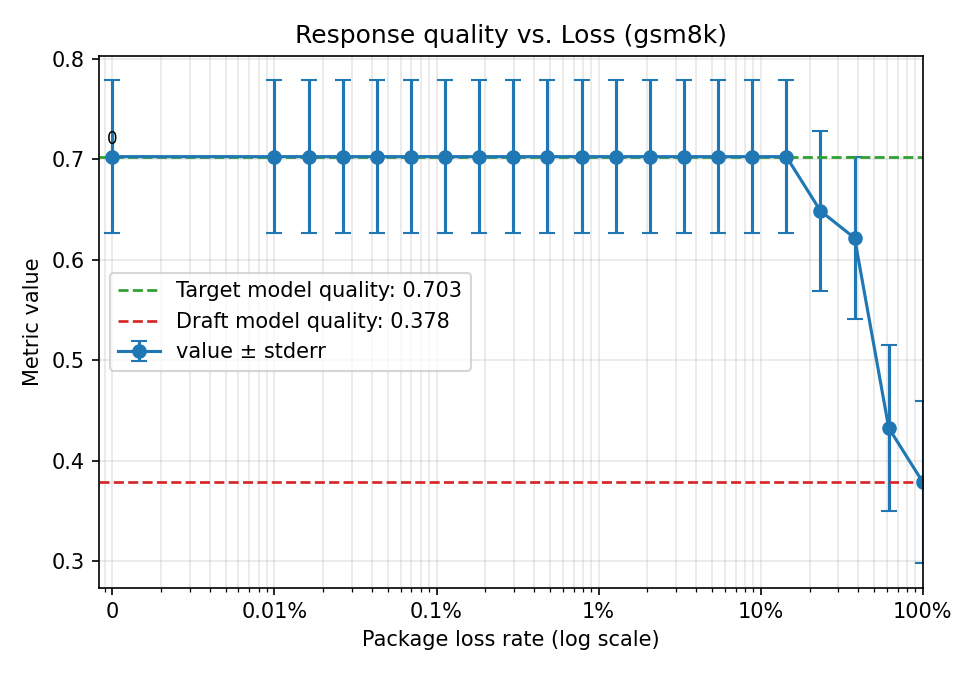}
    \caption{\textcolor{black}{The impact of packet loss on the quality of generated response.}}
    \Description{Line plot of response quality versus packet loss on gsm8k; quality matches the target model at low loss and declines as loss grows, approaching the draft-only level when loss is extreme.}
    \label{fig_packageloss_quality}
\end{figure}

\textcolor{black}{Additionally, we evaluated the response quality of the \acronym under varying packet loss rates in the prototype system. By sweeping the packet loss ratio from 0\% to 100\%, we assessed performance on GSM8K \cite{Cobbe2021GSM8K}, a curated benchmark of linguistically diverse, grade-school math word problems designed to evaluate multi-step quantitative reasoning in language models. As shown in Fig.\ref{fig_packageloss_quality}, the response quality remains consistent with that of the target model when the verification packet loss rate is below 10\%, which covers the most common real-world conditions. The quality gradually declines to the draft model level as the packet loss rate increases from 10\% to 100\%, indicating a loss of connection to the verification server. Based on the proposed retry-then-fallback policy, persistent packet loss triggers a fallback to the local draft model for token generation, leading to a degradation in output quality. These results demonstrate that under typical connectivity conditions, \acronym can effectively serve edge devices without compromising accuracy.}

Overall, the experimental evaluations underscore the significant advantages of \acronym in distributed inference scenarios, including throughput, capacity, and cost efficiency, showcasing insightful findings in the \acronym system to motivate more explorations in future work.

\section{Conclusion and Future Work} \label{sec:conclusion}
This position paper presented the \acronym, a novel distributed decoding framework designed for LLM deployment at the edge. Our extensive evaluation demonstrated that \acronym significantly improves system throughput, capacity, and cost efficiency compared to traditional centralized approaches. The integration of speculative local drafting and centralized verification establishes a balance of computational workload, making \acronym particularly suitable for bringing LLMs towards the edge of the network. The position paper highlights that the \acronym is more than a decoding enhancement—it opens the door to a more foundational and elastic approach to resource-aware LLM serving at the edge.

\textcolor{black}{In future work, we will explore optimizing key–value caching strategies, potentially leveraging recent advances in PagedAttention, to further reduce server-side verification cost.} Additionally, enhancing the adaptive capabilities of \acronym for dynamic environments, such as extending \acronym's applicability to multi-modal scenarios, will be another interesting topic to focus on. Lastly, network conditions and resource-aware verification strategy could further broaden its practical impact in complex edge computing landscapes.

\balance

\begin{acks}
This material is based on work supported by the National Science Foundation under Grants No. 2315851 and 2106634.
\end{acks}

\bibliographystyle{ACM-Reference-Format}
\bibliography{sample-base}


\end{document}